\documentclass[journal]{IEEEtran}
\usepackage{commath,graphicx,afterpage,subfigure,amsmath,amsfonts,amssymb,algorithm,algorithmic}
\usepackage{cite,multirow,bm,multicol,booktabs,threeparttable,extpfeil}

\begin{document}

\title{T-TIME: Test-Time Information Maximization Ensemble for Plug-and-Play BCIs}

\author{Siyang~Li, Ziwei~Wang, Hanbin Luo, Lieyun Ding, and Dongrui~Wu

\thanks{S.~Li, Z.~Wang and D.~Wu are with the Ministry of Education Key Laboratory of Image Processing and Intelligent Control, School of Artificial Intelligence and Automation, Huazhong University of Science and Technology, Wuhan 430074, China. They are also with the Shenzhen Huazhong University of Science and Technology Research Institute, Shenzhen, China. S.~Li is also with the Henan Key Laboratory of Brain Science and Brain Computer Interface Technology, School of Electrical and Information Engineering, Zhengzhou University.}
\thanks{H.~Luo and L.~Ding are with the School of Civil and Hydraulic Engineering, Huazhong University of Science and Technology, Wuhan 430074 China.}
\thanks{Corresponding Authors: Lieyun Ding (dly@hust.edu.cn) and Dongrui Wu (drwu@hust.edu.cn).}}

\maketitle

\begin{abstract}
\emph{Objective}: An electroencephalogram (EEG)-based brain-computer interface (BCI) enables direct communication between the human brain and a computer. Due to individual differences and non-stationarity of EEG signals, such BCIs usually require a subject-specific calibration session before each use, which is time-consuming and user-unfriendly. Transfer learning (TL) has been proposed to shorten or eliminate this calibration, but existing TL approaches mainly consider offline settings, where all unlabeled EEG trials from the new user are available. \emph{Methods}: This paper proposes Test-Time Information Maximization Ensemble (T-TIME) to accommodate the most challenging online TL scenario, where unlabeled EEG data from the new user arrive in a stream, and immediate classification is performed. T-TIME initializes multiple classifiers from the aligned source data. When an unlabeled test EEG trial arrives, T-TIME first predicts its labels using ensemble learning, and then updates each classifier by conditional entropy minimization and adaptive marginal distribution regularization. Our code is publicized.
\emph{Results}: Extensive experiments on three public motor imagery based BCI datasets demonstrated that T-TIME outperformed about 20 classical and state-of-the-art TL approaches.
\emph{Significance}: To our knowledge, this is the first work on test time adaptation for calibration-free EEG-based BCIs, making plug-and-play BCIs possible.
\end{abstract}


\section{Introduction}

A brain-computer interface (BCI)~\cite{Graimann2009} enables direct communication between the human brain and a computer by neural activities of the user. It can assist, augment, or even repair human cognitive or sensory–motor functions~\cite{Krucoff2016}. In addition to rehabilitation for disabled people, BCIs have also found applications for able-bodied users in device control, education, gaming, and so on~\cite{Erp2012}.

BCIs can be categorized into non-invasive, partially invasive, and invasive ones. Among them, non-invasive BCIs, which usually use electroencephalogram (EEG) as the input signal~\cite{Wu2022a}, are the most convenient. The three classic paradigms in EEG-based BCIs are event-related potentials~\cite{Sutton1965}, steady-state visual evoked potentials~\cite{Friman2007}, and motor imagery (MI)~\cite{Pfurtscheller2001}. In an MI-based BCI, the user imagines the movement of his/her body parts (e.g., left hand, right hand, both feet, or tongue), which modulates different areas of the motor cortex of the brain~\cite{Wu2022}. These imageries are then mapped into specific commands for external device control.

Despite various advantages, such as noninvasiveness and low cost, EEG demonstrates significant individual differences, i.e., for the same task, different users usually exhibit different EEG responses. The current solution requires a subject-specific calibration session before each use, which is time-consuming and user-unfriendly, hindering their broad real-world applications.

\begin{figure}[htbp]\centering
\subfigure[]{\label{fig:uda}   \includegraphics[width=.9\linewidth,clip]{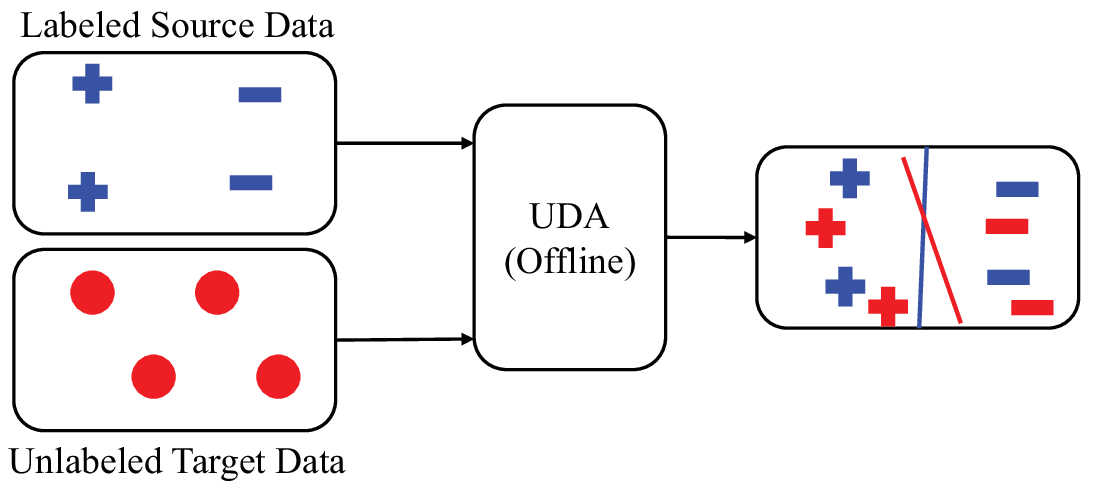}}
\subfigure[]{\label{fig:sfuda}    \includegraphics[width=.95\linewidth,clip]{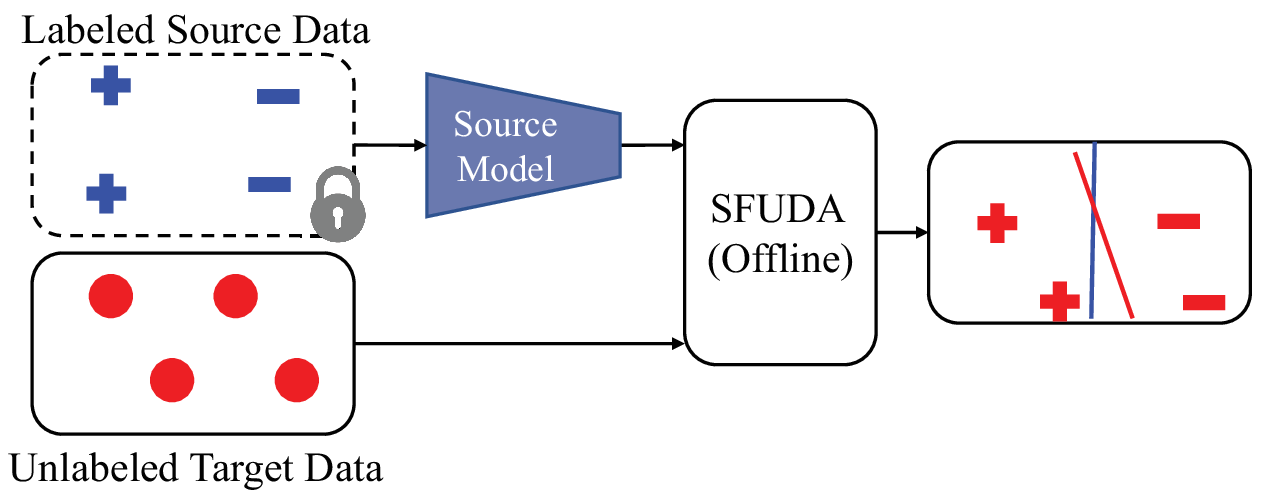}}
\subfigure[]{\label{fig:tta}   \includegraphics[width=\linewidth,clip]{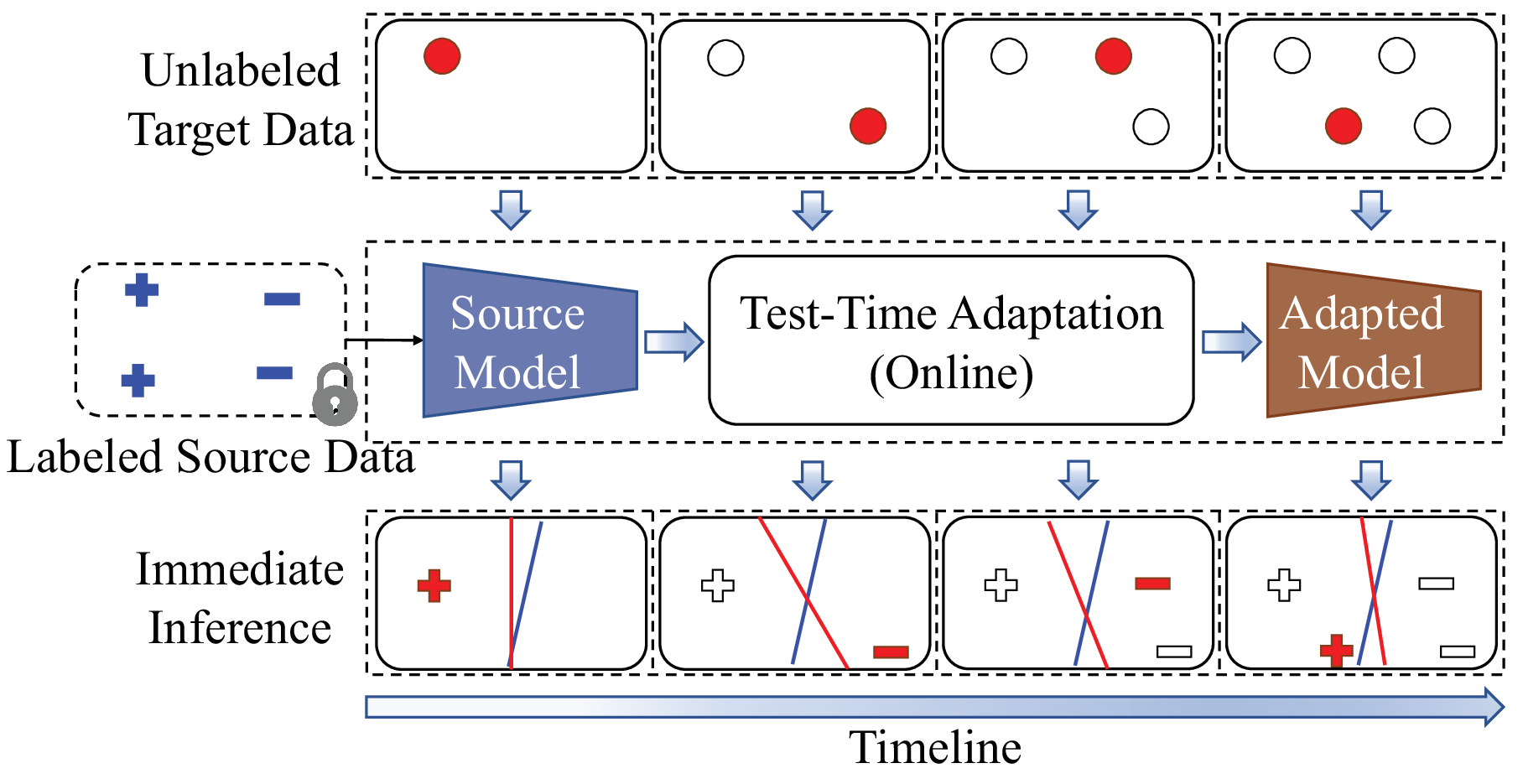}}
\caption{Three different TL settings, when the target domain is completely unlabeled. (a) Unsupervised domain adaptation (UDA); (b) source-free unsupervised domain adaptation (SFUDA); and, (c) test-time adaptation (TTA).} \label{fig:settings}
\end{figure}

Transfer learning (TL)~\cite{Wu2022, Wu2022a}, which utilizes knowledge from previous users (source domains) to facilitate the learning for a new user (target domain), is a promising solution to alleviate individual differences and hence to reduce or eliminate the calibration effort. There are three different TL scenarios when the target domain is completely unlabeled:
\begin{enumerate}
\item Unsupervised domain adaptation (UDA), where the labeled source data and the unlabeled target data are combined for offline learning.

\item Source-free UDA (SFUDA), where the source model, instead of the source data, is used in UDA, usually for source domain privacy protection.

\item Test-time adaptation (TTA), where the source model is updated online to the target domain, whose test samples are classified in real-time.
\end{enumerate}
A comparison of these three settings is shown in Fig.~\ref{fig:settings}.

Existing TL approaches mainly consider offline UDA or SFUDA settings, where all unlabeled target data are available. However, in real-world online TTA applications, such as MI-based wheelchair control, unlabeled EEG trials arrive in a stream, and immediate classification is required.

This paper proposes Test-Time Information Maximization Ensemble (T-TIME) to accommodate the most challenging online TTA scenario for calibration-free EEG-based BCIs. T-TIME initializes multiple classifiers from the aligned source data. When an unlabeled test EEG trial arrives, T-TIME first predicts its labels using ensemble learning, and then updates each classifier by conditional entropy minimization and adaptive marginal distribution regularization. Extensive experiments on three public MI-based BCI datasets demonstrated that T-TIME outperformed about 20 classical and state-of-the-art TL approaches. To our knowledge, this is the first study on TTA for plug-and-play EEG-based BCIs. Our code is also publicized.

The remainder of this paper is organized as follows: Section~\ref{sect:relatedwork} introduces related work. Section~\ref{sect:approach} proposes T-TIME. Section~\ref{sect:experiments} presents experimental results to demonstrate the performance of T-TIME. Finally, Section~\ref{sect:conclusions} draws conclusions and points out some future research directions.

\section{Related Works}\label{sect:relatedwork}

This section introduces related works on deep learning based TL.

\subsection{Unsupervised Domain Adaptation}

UDA~\cite{Wilson2020} combines the labeled source data and the unlabeled target data for offline learning. The domain discrepancies can be reduced from three perspectives:
\begin{enumerate}
\item Input space. CORrelation ALignment (CORAL)~\cite{Sun2016CORAL} aligns the second-order statistics of different input distributions with a linear transformation. For multi-channel EEG signals, Riemannian geometry-based approach~\cite{Zanini2018RA} and Euclidean alignment (EA)~\cite{He2020EA} have demonstrated outstanding performance.

\item Feature space. Deep adaptation network (DAN)~\cite{Long2015DAN} and joint adaptation network (JAN)~\cite{Long2017JAN} perform feature alignment by minimizing the maximum mean discrepancies. Domain-adversarial neural network (DANN)~\cite{Ganin2016DANN} and conditional domain adversarial network (CDAN)~\cite{Long2018CDAN} use adversarial training to reduce the feature discrepancies between the source and target domains.

\item Output space. Minimum class confusion (MCC)~\cite{Jin2020MCC} utilizes a weighted prediction entropy to minimize the class confusion. Liang~\emph{et al.}~\cite{Liang2021} introduced an auxiliary classifier to improve the quality of the target domain pseudo labels.
\end{enumerate}

\subsection{Source-free Unsupervised Domain Adaptation}

SFUDA~\cite{Fang2022} uses the source model, instead of the source data, for UDA. It is appropriate for source domain privacy protection, but generally more challenging to implement.

Source HypOthesis Transfer (SHOT)~\cite{Liang2022SHOT} uses information maximization and self-supervised learning for feature extraction in the target domain. To explicitly consider class-imbalance, imbalanced source-free domain adaptation (ISFDA)~\cite{Li2021ISFDA} uses intra-class tightening and inter-class separation to form better decision boundaries.

SFUDA has also been used in EEG-based BCIs for privacy protection~\cite{Xia2022}. Li \emph{et al.}~\cite{Li2022MDMAML} proposed a meta-learning strategy for multi-source model transfer. Zhang \emph{et al.} considered lightweight SFUDA~\cite{Zhang2022LSFT} and multi-source decentralized SFUDA~\cite{Zhang2022MSDT} for privacy-preserving BCIs.

\subsection{Test-time Adaptation}

Both UDA and SFUDA assume all unlabeled target data are available, i.e., they target at offline applications. On the contrast, TTA~\cite{Liang2023} focuses on online applications. In TTA, target data arrive in a stream, and a prediction must be immediately made for each coming sample. Thus, it is more challenging than UDA and SFUDA.

Target domain pseudo-labels~\cite{DongHyun2013PL} can be used for model adaptation in TTA. For example, Chen \emph{et al.}~\cite{Chen2022AdaContrast} combined contrastive learning with self-supervision on refined online pseudo-labels, and test-time template adjuster (T3A)~\cite{Iwasawa2021T3A} efficiently integrates few-shot learning with pseudo-labeling.

Entropy-based uncertainty reduction has also been employed in TTA. For example, test entropy minimization (Tent)~\cite{Wang2021TENT} performs TTA on normalization layers, and  marginal entropy minimization with one test point (MEMO)~\cite{Zhang2022MEMO} utilizes self-supervised augmentations for marginal entropy minimization.

The literature has also considered more challenging and complex scenarios that general TTA approaches may fail. Sharpness-aware and reliable entropy minimization (SAR)~\cite{Niu2023SAR} selects samples with smaller entropy losses and jointly minimizes the sharpness of the entropy and the entropy loss for a more reliable adaptation. Degradation-freE fuLly Test-time Adaptation (DELTA)~\cite{Zhao2023DELTA} uses dynamic online re-weighting (a momentum-updated class frequency) to weight each test sample in calculating the conditional entropy to cope with test-time class-imbalance.

Continual TTA handles test data flow with continually changing distribution. Continual test-time adaptation (CoTTA)~\cite{Wang2022CoTTA} utilizes pseudo-labels that are both weight-averaged and augmentation-averaged for adaptation in a teacher-student model.  Lange \emph{et al.}~\cite{DeLange2021CoPE} proposed a pseudo-prototypical proxy loss to encourage inter-class variance and reduce intra-class variance with a class-balanced data replay strategy for continual learning.

\section{Test-Time Information Maximization Ensemble (T-TIME)} \label{sect:approach}

We focus on closed set domain adaptation, i.e., the source and target domains have identical input and label spaces but different marginal and conditional probability distributions.

\subsection{Problem Setup}

Assume the $K$-class classification problem involves $L$ source subjects, and the $l$-th ($l=1,...,L$) source subject has $n_{s,l}$ labeled EEG trials $\{(X_{s,l}^i, y_{s,l}^i)\}_{i=1}^{n_{s,l}}$, where $X_{s,l}^i \in \mathbb{R}^{ch \times ts}$ is the $i$-th EEG trial ($ch$ is the number of EEG channels, and $ts$ the number of time samples), and $y_{s,l}^i$ the corresponding label. $n_t$ test EEG trials $\{X_t^i\}_{i=1}^{n_t}$ from the target subject arrive online sequentially, and the goal is to predict their labels $\{y_t^i\}_{i=1}^{n_t}$ in a completely unsupervised and online manner. More specifically, at test time $a$, only the source data and $\{X_t^i\}_{i=1}^a$ can be used to make the prediction $\hat{y}_t^a$ for $X_t^a$.

\begin{figure}[htpb]
\includegraphics[width=\linewidth,clip]{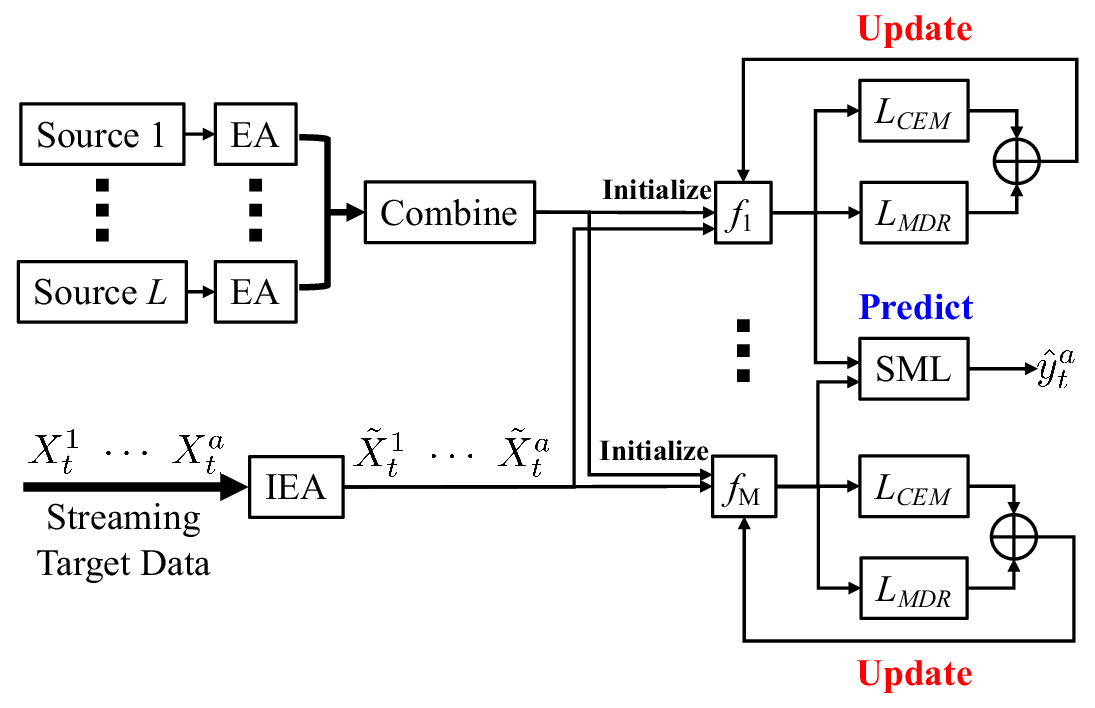}
\caption{Flowchart of the proposed T-TIME.} \label{fig:flowchart}
\end{figure}

Our proposed T-TIME, shown in Fig.~\ref{fig:flowchart}, uses the source model predictions on the test data to calibrate the target model, without using any labeled target data.

\subsection{Source Model Training}

We perform Euclidean alignment (EA)~\cite{He2020EA} on each source subject individually to reduce the individual differences, and then combine all aligned EEG trials from the $L$ source subjects into a single source domain to train $M$ source models.

For the $l$-th ($l=1,...,L$) source subject, EA computes the arithmetic mean of all covariance matrices of his/her EEG trials~\cite{He2020EA}:
\begin{align}
\bar{R}_{s,l} = \frac{1}{n_{s,l}} \sum_{i=1}^{n_{s,l}} X_{s,l}^i (X_{s,l}^i)^\top, \label{eq:EA-cov}
\end{align}
and then performs the alignment by
\begin{align}
\tilde{X}_{s,l}^i= \bar{R}_{s,l}^{-1/2} X_{s,l}^i, \quad i=1,...,n_{s,l}. \label{eq:EA-align}
\end{align}
All $\bigcup\limits_{l=1,...,L}\{\tilde{X}_{s,l}^i\}_{i=1}^{n_{s,l}}$ are then assembled into $\{\tilde{X}_s^i\}_{i=1}^{n_s}$ and viewed as a single source domain, where $n_s=\sum_{l=1}^L n_{s,l}$. The labels are accordingly assembled into $\{y_s^i\}_{i=1}^{n_s}$.

$M$ EEGNet~\cite{Lawhern2018EEGNet} models $\{f_m\}_{m=1}^M$, each with random initialization, can then be independently trained from $\{(\tilde{X}_s^i,y_s^i)\}_{i=1}^{n_s}$ by minimizing the classic cross-entropy loss.

\subsection{Incremental EA (IEA) on Target Data}

The target EEG trials arrive online one by one, so we perform incremental EA on them.

More specifically, when $X_t^a$ arrives, we first update:
\begin{align}
\bar{R}_t^{a}=\frac{1}a \sum_{i=1}^a X_t^i (X_t^i)^\top, \label{eq:IEA}
\end{align}
and then perform EA on $\{X_t^i\}_{i=1}^a$ using:
\begin{align}
\tilde{X}_t^i= {\left(\bar{R}_t^{a}\right)}^{-1/2} X_t^i, \quad i=1,...,a. \label{eq:EA-TTA}
\end{align}
$\{\tilde{X}_t^i\}_{i=1}^a$ then replace $\{X_t^i\}_{i=1}^a$ as the input to $\{f_m\}_{m=1}^M$ for classification.

\subsection{Target Label Prediction}  \label{subsect:Prediction}

The $M$ TTA models $\{f_m\}_{m=1}^M$ are initialized from the $M$ source models trained above, and updated after the arrival of each $X_t^i$, using the procedure introduced in the next subsection.

Assume $\{f_m\}_{m=1}^M$ have been updated up to $X_t^{a-1}$. When the next test EEG trial $X_t^a$ arrives, it is transformed into $\tilde{X}_t^a$ using incremental EA introduced in the previous subsection, input into each $f_m$ for classification, and then the $M$ probability vectors $\{f_m(\tilde{X}_t^a)\}_{m=1}^M$ are combined using ensemble learning to obtain the predicted label $\hat{y}_t^a$ :
\begin{enumerate}
\item When $a\le M$, the individual probability vectors $\{f_m(\tilde{X}_t^a)\}_{m=1}^M$ are averaged, and $\hat{y}_t^a$  is the class with the largest average probability.
\item When $a>M$, the spectral meta-learner (SML)~\cite{Parisi2014SML} is used to weighted average $\{f_m(\tilde{X}_t^a)\}_{m=1}^M$. SML constructs a meta-classifier, more accurate than most, if not all, of the individual classifiers, using only their predictions on the unlabeled test data.

The original SML was proposed for binary classification with 0/1 predictions. To accommodate the TTA setting, we use it for multi-class classification with continuous prediction probabilities.

Let $F_k(\tilde{X}_t^i) =  \left[\delta_k \left(f_1(\tilde{X}_t^i)\right); ...; \delta_k \left(f_M(\tilde{X}_t^i)\right)\right]^\top$ be the predicted probabilities of all $M$ models for class $k$, where $\delta_k$ is the $k$-th element of the softmax output of $f_m(\tilde{X}_t^i)$. SML first computes the sample covariance matrix, $Q_k \in \mathbb{R}^{M \times M}$, of the $M$ classifiers~\cite{Parisi2014SML}:
\begin{align}
Q_k = \frac{1}{a-1}  \sum_{i=1}^a &\left(F_k(\tilde{X}_t^i) - \mathbb{E}[F_k(\tilde{X}_t)]\right) \nonumber\\
&\cdot {\left(F_k(\tilde{X}_t^i) - \mathbb{E}[F_k(\tilde{X}_t)]\right)}^\top. \label{eq:SML-cov}
\end{align}
$Q_k$ approximates a rank-one matrix, and the entries of its principal eigenvector $\bm{v}_k$ are proportional to the balanced classification accuracies of the $M$ models~\cite{Parisi2014SML}. A meta-learner can then be constructed by a weighted combination of the $M$ prediction probabilities:
\begin{align}
\hat{y}_t^a = \arg\max_{k}  \sum_{m=1}^M \delta_k\left(f_m(\tilde{X}_t^a)\right) \cdot v_{k, m}, \label{eq:SML-ensemble}
\end{align}
where $v_{k, m}$ is the $m$-th element of $\bm{v}_{k}$.
\end{enumerate}

\subsection{Target Model Update}

The target models are updated using a sliding batch with size $B$. When there are not enough target data, i.e., $a<B$, the target models are fixed to be $\{f_m\}_{m=1}^M$. They are updated for every $X_t^a$ when $a\ge B$.

\begin{figure}[htpb]
\includegraphics[width=\linewidth,clip]{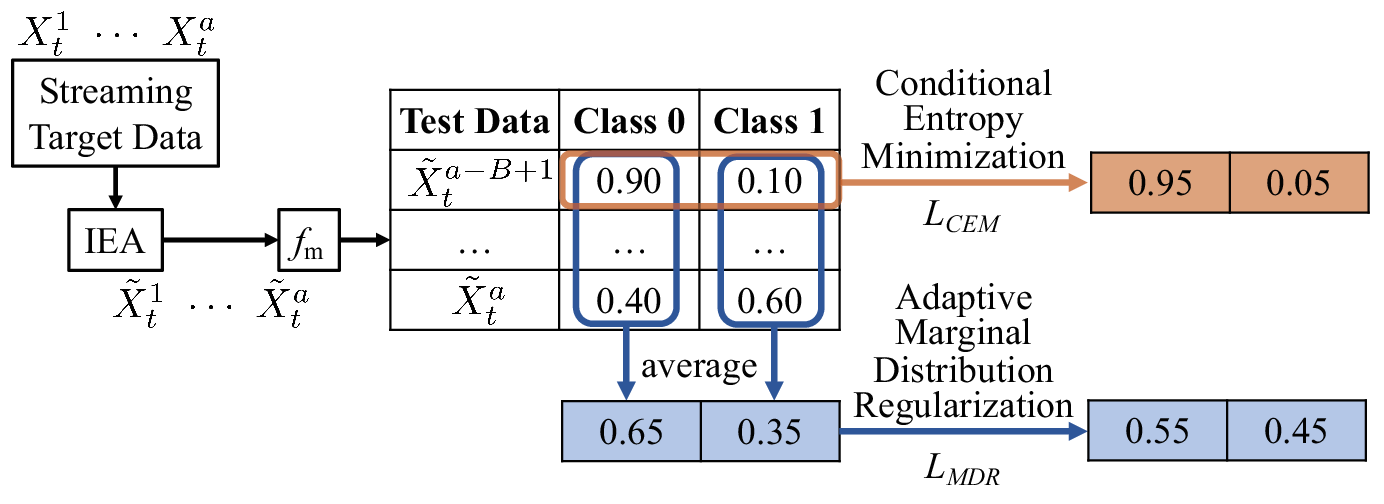}
\caption{Conditional entropy minimization and adaptive marginal distribution regularization in target model update.} \label{fig:reg}
\end{figure}

When $X_t^a$ arrives, $\{X_t^i\}_{i=a-B+1}^a$ are first transformed into $\{\tilde{X}_t^i\}_{i=a-B+1}^a$ through incremental EA, and then the loss function for updating $f_m$ is:
\begin{align}
\mathcal{L}_m =  &\mathcal{L}_{CEM} (f_m; \{\tilde{X}_t^i\}_{i=a-B+1}^a)\nonumber\\
& +  \mathcal{L}_{MDR} (f_m; \{\tilde{X}_t^i\}_{i=a-B+1}^a), \label{eq:update}
\end{align}
where $\mathcal{L}_{CEM}$ and $\mathcal{L}_{MDR}$ account for conditional entropy minimization and adaptive marginal distribution regularization, respectively, as illustrated in Fig.~\ref{fig:reg} and detailed next.

\subsubsection{Conditional Entropy Minimization}

Conditional entropy minimization suppresses the within-sample prediction uncertainty, i.e., it forces the adapted model to have high prediction confidence on the test data.

We fine-tune all parameters of $f_m$, instead of only the normalization layers~\cite{Wang2021TENT}, which is unstable and may fail under hard cases~\cite{Niu2023SAR}. Specifically, our loss function uses Shannon entropy to measure the conditional entropy of the predicted probabilities:
\begin{align}
 &\mathcal{L}_{CEM} (f_m;\{\tilde{X}_t^i\}_{i=a-B+1}^a)\nonumber\\
 =& - \frac{1}B\sum_{i=a-B+1}^a \sum_{k=1}^{K} \delta_k \left(\frac{f_m(\tilde{X}_t^i)}{T}\right)
 \cdot\log \delta_k \left(\frac{f_m(\tilde{X}_t^i)}{T}\right). \label{eq:CEM}
\end{align}
Note that temperature scaling~\cite{Guo2017} with factor $T$ is used to recalibrate the model's prediction confidence on the target data. $\mathcal{L}_{CEM} $ forces $f_m$ to have a high prediction probability towards one specific class on each test sample, resulting in clearer class boundaries in the target domain.

\subsubsection{Adaptive Marginal Distribution Regularization}

Only minimizing the conditional entropy above may result in two undesirable outcomes: 1) a trivial solution that all test data are classified into a single class; or, 2) a wrong classification becomes overly confident. For remedy, we further add within-batch marginal label distribution regularization.

Specifically, the average prediction probability $\bar{p}_k$ (also with temperature factor $T$) for class $k$ in the sliding batch is:
\begin{align}
\bar{p}_k = \frac{1}B\sum_{i=a-B+1}^a \delta_k \left(\frac{f_m(\tilde{X}_t^i)}{T}\right). \label{eq:avgpred}
\end{align}
The entropy sum of $\bar{p}_k$ evaluated on each target batch was used in~\cite{Liang2022SHOT} as a diversity loss for regularizing $\mathcal{L}_{CEM}$ in SFUDA, also known as the information maximization (IM) loss, assuming the target domain is class-balanced. Such an assumption may be true on simple offline tasks with a large batch size, but could lead to biased optimization towards the dominant class~\cite{Zhao2023DELTA} for online TTA with class-imbalance and small batch size~\cite{Niu2023SAR, Boudiaf2022LAME}.

Thus, we propose adaptive label marginal distribution regularization for class-imbalanced test batches. During adaptation, the target domain class-frequency is estimated using pseudo-labeling with confidence threshold on the test data. Specifically, the estimated target class-frequency $z_k$ for class $k$ at test time $a$ is:
\begin{align}
z_k =\left| \{ X_t^i \mid i \in [a-B+1, a],  \delta_k\left(f_m(\tilde{X}_t^i)\right) \geq \tau \} \right|, \label{eq:frequency}
\end{align}
where $\tau\in[0.5, 1.0)$ is a hyper-parameter of pseudo-labeling confidence threshold.

The average prediction probability $\bar{p}_k$ is recalibrated as:
\begin{align}
q_k = \frac{\bar{p}_k}{c+ z_k},\label{eq:calibpred}
\end{align}
where $c$ is a small integer to avoid dividing by zero, and also to make sure $q_k$ is neither too big nor too small at the beginning of the test phase.

The normalized $q_k$, i.e.,
\begin{align}
\hat{q}_k = \frac{q_k}{\sum_{i=1}^{K} q_i} ,\label{eq:balancepred}
\end{align}
is then used in adaptive marginal distribution regularization:
\begin{align}
 \mathcal{L}_{MDR} (f_m; \{\tilde{X}_t^i\}_{i=a-B+1}^a) = \sum_{k=1}^{K} \hat{q}_k \cdot \log \hat{q}_k. \label{eq:MDR}
\end{align}
This recalibration prevents falsely penalizing the model for skewed label distribution under observed class-imbalance.

\subsection{The Complete T-TIME Algorithm}

The pseudo-code of T-TIME is given in Algorithm~\ref{tab:alg}. Note that $\{f_m\}_{m=1}^M$ are independently initialized and updated in both training and adaptation, consistent with the conditional independence assumption of SML~\cite{Parisi2014SML}. Note also that for each arriving $X_t^a$, its prediction is first made, and then $\{f_m\}_{m=1}^M$ are updated. The implication for deploying the algorithm in real-world BCI applications is that the algorithm would first make target label prediction, and then conduct target model update in parallel with downstream task execution. The pre-inference computation time only involves model inference, which should be fast. The post-inference computation time is slightly longer but still compatible with current BCI applications. More discussions on the computational cost will be given in Section~\ref{sect:comptime}.

The three main components of T-TIME consider information from three different granularity:
\begin{enumerate}
\item Conditional entropy minimization considers each test EEG trial in the sliding batch individually.
\item Adaptive marginal distribution regularization views the batch as a whole to accommodate class-imbalance.
\item SML aggregates different classifiers, which have incorporated information from both the source domain and all available target data.
\end{enumerate}

\begin{algorithm}[htpb]
 \caption{Test-Time Information Maximization Ensemble (T-TIME).}
 \label{tab:alg}
 \begin{algorithmic}
\REQUIRE $L$ source domains, each with labeled data $\{(X_{s,l}^i, y_{s,l}^i)\}_{i=1}^{n_{s,l}}$, $l=1,...,L$;\\
 Streaming target data $\{X_t^i\}_{i=1}^a$;\\
 $M$, the number of base classifiers;\\
 $B$, the sliding batch size;\\
 $T$, the temperature rescaling factor in (\ref{eq:CEM}) and (\ref{eq:avgpred});\\
 $\tau$, the confidence threshold in (\ref{eq:frequency});\\
 $c$, the small integer in (\ref{eq:calibpred});\\
\ENSURE The classification  $\hat{y}_t^a$ for $X_t^a$.
\STATE \emph{// Source Model Training}
\FOR{$l=1:L$}
\STATE Perform EA on $\{X_{s,l}^i\}_{i=1}^{n_{s,l}}$ by (\ref{eq:EA-cov}) and (\ref{eq:EA-align}) to obtain $\{\tilde{X}_{s,l}^i\}_{i=1}^{n_{s,l}}$;
\ENDFOR
\STATE Assemble all $\bigcup\limits_{l=1,...,L}\{\tilde{X}_{s,l}^i\}_{i=1}^{n_{s,l}}$  into $\{\tilde{X}_s^i\}_{i=1}^{n_s}$, and construct the corresponding $\{y_s^i\}_{i=1}^{n_s}$.
\STATE Train $M$ models ${\{f_m\}}_{m=1}^M$ independently on $\{(\tilde{X}_{s}^i, y_{s}^i)\}_{i=1}^{n_s}$;
\STATE \emph{// Target Lata Prediction}
  \STATE Initialize $z_k=0$, $k=1,...,K$;
  \STATE Perform incremental EA on $\{X_t^i\}_{i=1}^a$ by (\ref{eq:IEA}) and (\ref{eq:EA-TTA}) to obtain $\{\tilde{X}_t^i\}_{i=1}^a$;
    \IF{$a  \ge M$}
   \STATE Calculate $Q_k$ on $\{\tilde{X}_t^i\}_{i=1}^a$ by (\ref{eq:SML-cov});
  \STATE Calculate $\bm{v}_k$ of $Q_k$;
  \STATE Calculate $\hat{y}_t^a$ for $\tilde{X}_t^a$ by (\ref{eq:SML-ensemble});
    \ELSE
    \STATE Calculate $\hat{y}_t^a$, which is $k$ that maximizes the averaged $F_k(\tilde{X}_t^a)$;
      \ENDIF
  \STATE\emph{// Target Model Update}
  \IF{$a\ge B$}
      \FOR{$m=1:M$}
    \STATE Calculate $\mathcal{L}_{CEM}$ on ${\{\tilde{X}_t^i\}}_{i={(a-B+1)}}^a$ by (\ref{eq:CEM});
    \STATE Calculate $\mathcal{L}_{MDR}$ on ${\{\tilde{X}_t^i\}}_{i={(a-B+1)}}^a$ by (\ref{eq:avgpred})-(\ref{eq:MDR});
      \STATE Calculate $\mathcal{L}$ by (\ref{eq:update}) and update $f_m$;
    \ENDFOR
    \ENDIF
 \end{algorithmic}
 \end{algorithm}

\section{Experiments}\label{sect:experiments}

Extensive experiments were performed to validate the superior performance of T-TIME.

\subsection{Datasets}\label{sect:dataset}

Three EEG-based MI benchmark datasets from MOABB~\cite{Jayaram2018} were used in our experiments. Their characteristics are summarized in Table~\ref{tab:datasets}. For BNCI2014001, only two classes (left/right hand imaginations) were used. For all three datasets, only data from the first session of each subject were used for training and test. The standard preprocessing steps in MOABB, including notch filtering, band-pass filtering, etc., were used to ensure the reproducibility.

\begin{table*}[htpb]  \center
\caption{Summary of the three MI EEG datasets.}  \label{tab:datasets}
\begin{tabular}{c|c|c|c|c|c|c|c}
\toprule
\multirow{2}{*}{Dataset} & Number of & Number of & Sampling & Trial Length & Number of  & Number of Trials & Types of \\
 & Subjects & Channels & Rate (Hz) & (seconds) & Sessions & in the First Session & Imaginations \\
\midrule
BNCI2014001 & 9 & 22 & 250 & 4 & 2 & 144 & left hand, right hand \\
BNCI2014002 & 14 & 15 & 512 & 5 & 2 & 100 & right hand, both feet \\
BNCI2015001 & 12 & 13 & 512 & 5 & 2 or 3 & 200 & right hand, both feet \\
\bottomrule
\end{tabular}
\end{table*}

\subsection{Algorithms} \label{sect:algorithms}

We compared T-TIME with about 20 classical and state-of-the-art algorithms, including traditional machine learning approaches, end-to-end deep neural networks, and UDA/SFUDA/TTA approaches:
\begin{enumerate}
\item CSP~\cite{Blankertz2008}, which first used the source domain labeled data to design Common Spatial Pattern (CSP) filters, and then performed feature extraction and linear discriminant analysis classification.
\item EEGNet~\cite{Lawhern2018EEGNet}, a popular end-to-end Convolutional Neural Network (CNN) for EEG signal decoding. We used its latest version-4 implementation with a default cross-subject setting, which has two blocks of CNN layers and a fully-connected classification layer.
\item UDA approaches, including DAN~\cite{Long2015DAN}, JAN~\cite{Long2017JAN}, DANN~\cite{Ganin2016DANN}, CDAN+E~\cite{Long2018CDAN}, MDD~\cite{Zhang2019MDD}, and MCC~\cite{Jin2020MCC}, all with EEGNet backbone.
\item SFUDA approaches, including SHOT~\cite{Liang2022SHOT}, SHOT-IM~\cite{Liang2022SHOT}, and ISFDA~\cite{Li2021ISFDA}, all with EEGNet backbone. Source models in SFUDA were the same as the EEGNet baseline.
\item TTA approaches, including BN-adapt~\cite{Schneider2020BNadapt}, Tent~\cite{Wang2021TENT}, PL~\cite{DongHyun2013PL}, T3A~\cite{Iwasawa2021T3A}, CoTTA~\cite{Wang2022CoTTA}, SAR~\cite{Niu2023SAR}, DELTA~\cite{Zhao2023DELTA}, and T-TIME, all with EEGNet backbone. All TTA approaches used sliding batches in optimization. Source models in TTA were also the same as the EEGNet baseline.
\end{enumerate}

For a fair comparison, the performance of T-TIME with and without the ensemble is separately listed. T-TIME (5) refers to the complete configuration with SML (Section~\ref{subsect:Prediction}) using 5 independently trained EEGNets with different random seeds.

EA~\cite{He2020EA} was applied before all approaches except otherwise explicitly stated. Each source subject was aligned independently using (\ref{eq:EA-cov}) and (\ref{eq:EA-align}). For the target subject, offline algorithms performed EA using all target data, whereas online algorithms performed incremental using (\ref{eq:IEA}) and (\ref{eq:EA-TTA}).

Leave-one-subject-out cross-validation was considered, i.e., each subject in the corresponding dataset was treated as the unlabeled test subject once, with all remaining subjects combined as the source domain. The source models were trained using the combined source domain, and the performance was tested on the left-out target domain (the test data arrived sequentially one-by-one in the online setting). All experiments on deep neural networks were repeated five times to accommodate randomness. The average results on each subject, and the entire dataset, were reported.

The source EEGNet models were trained with batch size 32, learning rate $10^{-3}$, and 100 epochs using the Adam optimizer. UDA used the same settings and batch size for the target domain. SFUDA and TTA used the same learning rate and optimizer, but test batch size 8. Temperature scaling factor $T=2$ was used to recalibrate the target model's prediction confidence. More discussions on the hyperparameters of T-TIME are given in Section~\ref{sect:ablation}. Other hyperparameters of the baselines were set according to the recommendations in their original publications.

All algorithms were implemented in PyTorch, and the source code is available on GitHub\footnote{https://github.com/sylyoung/DeepTransferEEG}.

\subsection{Classification Accuracies on Balanced Classes} \label{sect:balanced}

This subsection considers the simplest setting, i.e., the test domain is class-balanced. The classification accuracies are shown in Tables~\ref{tab:BNCI2014001}-\ref{tab:BNCI2015001} for the three datasets, respectively. Observe that:

\begin{table*}[!h]     \centering
       \caption{Cross-subject classification accuracies  (\%) on BNCI2014001. The best accuracies in offline TL are marked with *. The best accuracies in online TL are marked in bold, and the second best by an underline.}  \label{tab:BNCI2014001}
    \begin{tabular}{c|c|c|c|c|c|c|c|c|c|c|c}   \toprule
        Setting & Approach & S0 & S1 & S2 & S3 & S4 & S5 & S6 & S7 & S8 & Avg. \\         \midrule
        \multirow{2}{*}{Baselines}  & CSP (w/o EA)  & 82.64 & 50.69 & 69.44 & 66.67 & 47.22 & 62.50 & 71.53 & 88.19 & 68.06 & 67.44 \\
        & EEGNet (w/o EA) & 73.61 & 55.69 & 81.67 & 64.44 & 51.25 & 75.00 & 58.19 & 90.28 & 81.53& 70.19$_{\pm1.87}$ \\
        \midrule
         \multirow{10}{*}{Offline TL}  & CSP & 83.33 & 52.08 & 97.92 & 75.00* & 56.25 & 67.36 & 72.22* & 88.19 & 71.53& 73.77 \\
        & EEGNet & 83.19 & 60.28 & 92.08 & 67.92 & 57.22 & 72.50 & 64.86 & 86.11 & 79.44& 73.73$_{\pm1.11}$ \\
        ~ & DAN & 76.67 & 63.89* & 94.44 & 70.42 & 59.31 & 75.69 & 63.75 & 84.72 & 80.83& 74.41$_{\pm1.05}$ \\
        ~ & JAN & 81.94 & 63.89* & 90.97 & 71.94 & 60.56 & 72.64 & 68.33 & 83.89 & 83.06& 75.25$_{\pm1.34}$ \\
        ~ & DANN & 82.64 & 60.97 & 91.39 & 68.89 & 63.47* & 79.31* & 65.42 & 84.31 & 82.36& 75.42$_{\pm0.79}$ \\
        ~ & CDAN-E & 80.14 & 63.33 & 92.36 & 71.11 & 57.92 & 74.03 & 68.33 & 87.78 & 87.36 & 75.82$_{\pm0.66}$\\
        ~ & MDD & 79.31 & 57.64 & 94.44 & 70.42 & 57.92 & 73.33 & 65.97 & 84.03 & 86.25& 74.37$_{\pm1.50}$ \\
        ~ & MCC & 86.81* & 62.36 & 98.47* & 73.19 & 58.61 & 72.64 & 66.94 & 94.17* & 96.39*& 78.84*$_{\pm0.82}$ \\
        ~ & SHOT & 81.39 & 61.25 & 93.47 & 64.31 & 60.97 & 75.28 & 65.56 & 85.14 & 86.94& 74.92$_{\pm0.74}$ \\
        ~ & SHOT-IM & 84.44 & 63.33 & 94.31 & 70.83 & 61.67 & 75.28 & 70.42 & 87.64 & 86.81& 77.19$_{\pm1.41}$ \\
        \midrule
        \multirow{11}{*}{Online TL}  & CSP & 80.56 & 53.47 & \underline{96.53} & \textbf{72.22} & 54.86 & 63.89 & \textbf{72.92} & 88.19 & 72.22 & 72.76\\
        ~ & EEGNet & 82.22 & \underline{60.56} & 92.50 & 67.78 & 56.39 & 72.64 & 64.17 & 85.28 & 80.14 & 73.52$_{\pm1.14}$\\
        ~ & BN-adapt & 80.83 & 59.72 & 92.78 & 69.44 & 57.64 & 72.08 & 67.08 & 84.86 & 86.39 & 74.54$_{\pm1.70}$\\
        ~ & Tent & 78.89 & 58.75 & 92.92 & 69.03 & 57.22 & 72.36 & 67.78 & 85.28 & 86.67& 74.32$_{\pm1.43}$ \\
        ~ & PL & 78.75 & 57.64 & 94.44 & 66.53 & \underline{59.44} & 72.36 & 69.86 & 88.61 & 88.33& 75.11$_{\pm0.89}$ \\
        ~ & T3A & 82.50 & 51.53 & 93.19 & 55.28 & 49.72 & 58.06 & 57.78 & 84.17 & 81.81& 68.23$_{\pm1.19}$ \\
        ~ & CoTTA & 75.69 & 59.31 & 91.94 & 66.81 & 55.56 & 71.67 & 62.78 & 84.44 & 80.14& 72.04$_{\pm0.56}$ \\
        ~ & SAR & 83.47 & 57.50 & 95.28 & 67.08 & 54.72 & 71.53 & 63.19 & 89.72 & 90.69& 74.80$_{\pm0.48}$ \\
        ~ & T-TIME & \underline{84.03} & \underline{60.56} & 95.69 & 67.64 & 57.22 & \underline{73.33} & 67.22 & \underline{91.25} & \underline{90.97} & \underline{76.44}$_{\pm0.55}$ \\
        ~ & T-TIME (5) & \textbf{84.93} & \textbf{63.54} & \textbf{96.88} & \underline{70.83} & \textbf{62.78} & \textbf{77.22} & \underline{71.81} & \textbf{92.22} & \textbf{93.47}& \textbf{79.30}$_{\pm0.82}$\\         \bottomrule
    \end{tabular}
\end{table*}

\begin{table*}[!h]
\addtolength{\tabcolsep}{-4pt}     \centering
       \caption{Cross-subject classification accuracies  (\%) on BNCI2014002. The best accuracies in offline TL are marked with *. The best accuracies in online TL are marked in bold, and the second best by an underline.}  \label{tab:BNCI2014002}
    \begin{tabular}{c|c|c|c|c|c|c|c|c|c|c|c|c|c|c|c|c}   \toprule
        Setting & Approach & S0 & S1 & S2 & S3 & S4 & S5 & S6 & S7 & S8 & S9 & S10 & S11 & S12 & S13& Avg. \\          \midrule
         \multirow{2}{*}{Baselines} & CSP (w/o EA) & 57.00 & 75.00 & 95.00 & 72.00 & 54.00 & 64.00 & 68.00 & 55.00 & 75.00 & 71.00 & 50.00 & 56.00 & 50.00 & 42.00 & 63.14 \\
        ~ & EEGNet (w/o EA) & 50.00 & 53.00 & 51.60 & 51.60 & 50.80 & 50.00 & 63.20 & 50.80 & 91.60 & 50.00 & 50.00 & 50.00 & 51.00 & 50.00& 54.54$_{\pm2.31}$ \\                 \midrule
        \multirow{10}{*}{Offline TL}  & CSP & 62.00 & 82.00 & 98.00 & 76.00 & 79.00 & 70.00 & 84.00 & 67.00 & 94.00* & 72.00 & 68.00 & 63.00 & 59.00 & 44.00 & 72.71\\
        ~ & EEGNet & 65.00 & 80.00 & 83.00 & 80.20 & 74.20 & 68.20 & 88.80 & 54.60 & 91.20 & 75.00 & 81.00 & 72.00 & 59.80 & 51.40 & 73.17$_{\pm0.59}$\\
        ~ & DAN & 67.80 & 79.40 & 87.00 & 82.40 & 74.80 & 67.80 & 85.40 & 64.80 & 89.80 & 75.20 & 80.80 & 72.60 & 56.20 & 50.80& 73.91$_{\pm0.60}$ \\
        ~ & JAN & 71.20 & 78.20 & 89.20 & 87.80 & 83.80* & 67.20 & 84.80 & 69.40 & 86.80 & 75.80 & 81.80 & 74.40 & 53.80 & 54.40& 75.61$_{\pm1.36}$ \\
        ~ & DANN & 68.00 & 80.60 & 89.20 & 72.80 & 77.40 & 63.60 & 87.20 & 61.00 & 87.60 & 75.00 & 82.80 & 73.60 & 59.00 & 56.80*& 73.90$_{\pm0.86}$ \\
        ~ & CDAN-E & 74.60* & 81.60 & 94.60 & 85.00 & 81.80 & 62.60 & 88.20 & 66.20 & 88.60 & 75.20 & 83.80 & 75.80 & 57.80 & 53.60& 76.39$_{\pm0.53}$ \\
        ~ & MDD & 68.60 & 80.20 & 85.00 & 81.80 & 77.00 & 67.00 & 88.80 & 61.40 & 91.20 & 73.80 & 82.80 & 76.00 & 63.60* & 54.40 & 75.11$_{\pm0.31}$ \\
        ~ & MCC & 70.80 & 82.80* & 99.00* & 94.60* & 83.80* & 72.60* & 89.00 & 79.00* & 89.60 & 79.80* & 87.60* & 82.80* & 56.80 & 53.00& 80.09*$_{\pm0.96}$ \\
        ~ & SHOT & 65.40 & 79.40 & 85.00 & 82.80 & 75.40 & 65.20 & 83.00 & 63.40 & 87.80 & 72.60 & 82.40 & 73.20 & 59.00 & 50.40& 73.21$_{\pm1.00}$ \\
        ~ & SHOT-IM & 73.40 & 80.40 & 94.00 & 84.40 & 81.80 & 70.60 & 90.80* & 65.60 & 88.00 & 76.00 & 84.20 & 74.40 & 63.60* & 50.80 & 77.00$_{\pm0.70}$\\                \midrule
         \multirow{11}{*}{Online TL} & CSP & 65.00 & 81.00 & \textbf{96.00} & 75.00 & 80.00 & 72.00 & 88.00 & 64.00 & \underline{90.00} & 75.00 & 73.00 & 66.00 & 58.00 & 41.00& 73.14 \\
        ~ & EEGNet & 65.80 & 79.80 & 82.60 & 78.60 & 70.20 & 67.60 & 87.80 & 55.20 & 89.60 & 75.20 & 79.20 & 72.60 & 58.40 & 54.00& 72.61$_{\pm0.53}$ \\
        ~ & BN-adapt & 67.60 & 77.80 & 83.20 & 84.60 & 72.00 & 68.60 & 84.20 & 61.60 & 88.20 & 74.20 & 80.60 & 68.80 & 56.00 & 55.60 & 73.07$_{\pm0.60}$\\
        ~ & Tent & 65.20 & 78.60 & 82.80 & 83.60 & 69.20 & 68.20 & 83.40 & 61.20 & 87.00 & 73.20 & 81.40 & 69.20 & 54.80 & 54.20& 72.29$_{\pm0.74}$ \\
        ~ & PL & 69.20 & 80.20 & 89.20 & 90.00 & 78.60 & 70.00 & 84.80 & 61.40 & 88.60 & 77.60 & 82.20 & 71.20 & 58.40 & \textbf{56.20}& 75.54$_{\pm0.34}$ \\
        ~ & T3A & 51.00 & 62.00 & 77.20 & 67.80 & 52.60 & 55.40 & 79.40 & 49.80 & \textbf{94.00} & 56.80 & 79.60 & 67.20 & 48.20 & 50.40 & 63.67$_{\pm3.01}$\\
        ~ & CoTTA & 65.40 & 76.80 & 83.00 & 79.60 & 72.40 & 67.60 & 87.80 & 59.60 & 87.60 & 74.80 & 77.80 & 71.40 & 56.80 & 52.00 & 72.33$_{\pm0.78}$\\
        ~ & SAR & \underline{70.60} & 80.80 & 93.40 & 91.80 & \underline{84.60} & 74.20 & \textbf{89.60} & \underline{66.80} & 86.20 & 78.20 & \textbf{84.00} & 76.60 & 55.20 & 53.40 & 77.53$_{\pm0.45}$ \\
        ~ & T-TIME & 70.00 & \underline{82.60} & 93.60 & 91.80 & 83.60 & \textbf{75.20} & \underline{89.40} & 65.60 & 87.00 & 78.80 & 82.80 & \underline{80.80} & \underline{59.80} & 54.60& \underline{78.26}$_{\pm0.50}$ \\
        ~ & T-TIME (5) & \textbf{74.80} & \textbf{83.50} & \underline{94.20} & \textbf{93.30} & \textbf{86.30} & \underline{74.40} & 89.10 & \textbf{69.50} & 87.70 & \underline{79.00} & \underline{83.60} & \textbf{83.50} & \textbf{63.00} & \underline{55.40}& \textbf{79.81}$_{\pm0.49}$ \\         \bottomrule
    \end{tabular}
\end{table*}

\begin{table*}[!h]
\addtolength{\tabcolsep}{-2pt}     \centering
       \caption{Cross-subject classification accuracies  (\%) on BNCI2015001. The best accuracies in offline TL are marked with *. The best accuracies in online TL are marked in bold, and the second best by an underline.}  \label{tab:BNCI2015001}
    \begin{tabular}{c|c|c|c|c|c|c|c|c|c|c|c|c|c|c}   \toprule
        Setting & Approach & S0 & S1 & S2 & S3 & S4 & S5 & S6 & S7 & S8 & S9 & S10 & S11& Avg. \\           \midrule
         \multirow{2}{*}{Baselines} & CSP (w/o EA) & 53.00 & 67.50 & 50.00 & 73.00 & 77.50 & 51.50 & 49.00 & 50.00 & 57.50 & 50.50 & 50.00 & 51.50 & 56.75\\
        ~ & EEGNet (w/o EA) & 52.80 & 84.00 & 53.00 & 84.80 & 71.70 & 66.00 & 68.50 & 65.70 & 64.00 & 57.90 & 50.60 & 54.90 & 64.49$_{\pm0.94}$\\                   \midrule
         \multirow{10}{*}{Offline TL} & CSP & 93.50 & 93.50 & 86.50 & 85.00 & 79.00 & 62.00 & 65.00 & 59.00 & 59.50 & 65.00 & 59.50 & 56.50* & 72.00\\
        ~ & EEGNet & 91.50 & 95.00 & 75.70 & 85.90 & 81.30 & 68.60 & 65.20 & 64.30 & 63.00 & 66.50 & 57.50 & 55.20& 72.48$_{\pm0.52}$ \\
        ~ & DAN & 94.20 & 93.60 & 86.70 & 86.00 & 82.00 & 67.00 & 68.40 & 63.50 & 63.70 & 69.10 & 59.80 & 55.60& 74.13$_{\pm0.76}$ \\
        ~ & JAN & 90.70 & 90.20 & 88.50 & 84.00 & 86.00 & 66.20 & 71.70 & 65.90 & 64.80 & 67.70 & 62.60 & 55.70& 74.50$_{\pm0.35}$ \\
        ~ & DANN & 94.00 & 94.20 & 83.90 & 84.90 & 87.00 & 69.30 & 72.10 & 64.50 & 63.10 & 64.90 & 52.10 & 50.80& 73.40$_{\pm0.54}$ \\
        ~ & CDAN-E & 95.20 & 94.50 & 90.70 & 85.70 & 86.90 & 69.20 & 82.90* & 64.80 & 66.10 & 66.90 & 61.20 & 53.50& 76.47$_{\pm0.69}$ \\
        ~ & MDD & 96.70 & 95.00 & 86.20 & 87.60 & 85.00 & 68.70 & 70.30 & 68.70* & 64.30 & 67.80 & 62.50 & 55.80& 75.72$_{\pm1.34}$ \\
        ~ & MCC & 97.90* & 95.40* & 95.00* & 89.80* & 94.30* & 67.30 & 81.30 & 65.90 & 68.90* & 69.50* & 68.70* & 52.20 & 78.85*$_{\pm0.41}$\\
        ~ & SHOT & 94.10 & 94.20 & 87.90 & 84.70 & 81.80 & 72.20* & 69.40 & 65.00 & 63.60 & 66.70 & 62.40 & 55.40& 74.78$_{\pm0.69}$ \\
        ~ & SHOT-IM & 96.60 & 94.30 & 90.80 & 86.80 & 91.20 & 71.30 & 72.90 & 67.40 & 65.00 & 69.20 & 58.60 & 54.60& 76.56$_{\pm0.74}$ \\                  \midrule
        \multirow{11}{*}{Online TL} & CSP & 96.50 & 95.00 & 86.50 & 85.50 & 81.50 & 64.00 & 69.50 & 54.50 & 57.50 & 62.50 & 57.00 & \underline{56.50} & 72.21\\
        ~ & EEGNet & 88.80 & \textbf{95.50} & 72.40 & 85.80 & 79.60 & \underline{69.80} & 66.00 & 62.30 & 64.50 & 65.90 & 54.10 & 55.50& 71.68$_{\pm0.72}$ \\
        ~ & BN-adapt & 96.00 & 94.60 & 90.50 & 84.20 & 82.80 & \textbf{71.00} & 69.10 & 62.70 & 62.90 & 66.10 & 61.30 & 54.20 & 74.62$_{\pm0.80}$\\
        ~ & Tent & 95.40 & 92.60 & 89.80 & 81.40 & 77.60 & 65.90 & 67.80 & 60.10 & 58.20 & 62.70 & 59.00 & 54.00& 72.04$_{\pm0.88}$ \\
        ~ & PL & 97.20 & 94.20 & 92.60 & 87.20 & 86.50 & 68.80 & 72.70 & 55.90 & 60.50 & 66.90 & \underline{61.70} & 54.40 & 74.88$_{\pm0.52}$\\
        ~ & T3A & 75.40 & 73.60 & 69.40 & 66.60 & 71.20 & 68.00 & 67.90 & 59.60 & 58.70 & 55.00 & \textbf{62.90} & 50.20 & 64.88$_{\pm2.90}$\\
        ~ & CoTTA & 93.10 & 93.70 & 85.90 & 81.40 & 75.90 & 67.90 & 65.80 & 57.20 & 62.30 & \textbf{67.70} & 60.00 & 54.80 & 72.14$_{\pm1.31}$\\
        ~ & SAR & 96.70 & 95.30 & 94.10 & 88.40 & 88.00 & 68.90 & 81.50 & 55.20 & 56.00 & 64.80 & 54.40 & 52.40 & 74.64$_{\pm1.15}$ \\
        ~ & T-TIME & \underline{97.50} & \underline{95.40} & \underline{94.80} & \textbf{89.10} & \underline{89.30} & 66.90 & \underline{83.60} & \textbf{68.80} & \textbf{66.60} & 65.20 & 59.60 & 56.20& \underline{77.75}$_{\pm0.67}$ \\
        ~ & T-TIME (5) & \textbf{97.60} & 94.40 & \underline{94.80} & \underline{88.85} & \textbf{90.95} & 67.40 & \textbf{85.55} & \underline{68.20} & \underline{65.80} & \underline{67.20} & 60.60 & \textbf{57.00} & \textbf{78.20}$_{\pm0.20}$\\             \bottomrule
    \end{tabular}
\end{table*}

\begin{enumerate}
\item The baselines without EA and TL did not work well for a new subject.

\item EA significantly improved the classification performance of CSP and EEGNet in both offline and online TL, and hence should be an essential data pre-processing step in TL.

\item Offline TL algorithms generally outperformed their online counterparts, which is intuitive, since offline TL had access to all test data.

\item End-to-end deep neural networks outperformed manual feature extraction approaches.

\item Overly complicated algorithms, e.g., CoTTA, deteriorated the classification performance, possibly due to inadequate training data. T3A used a fixed source model, and its performance indicated that building class prototypes with high dimensionality is difficult. BN-adapt and Tent only updated the batch-normalization layers, which seemed insufficient for our applications.

\item Sophisticated entropy-based approaches focusing on output-space TL, i.e., MCC, SHOT-IM, SAR and T-TIME, generally performed well. MCC had the best overall performance in offline TL.

\item Our proposed T-TIME performed the best among all online TL algorithms, and its performance was comparable with the best offline TL approach.
\end{enumerate}

\subsection{Classification Performance Under Class-Imbalance} \label{sect:imbalanced}

To simulate test-time class-imbalance, half of the Class 1 samples were randomly removed from the target data, leading to a class-imbalance ratio of 2:1. The source data were still balanced.

The area under the receiver operating characteristic curve (AUC) was used as the performance measure on the class-imbalanced data. The pseudo-labeling threshold $\tau=0.7$ and $c=4$ (half the test batch size) were used in adaptive marginal distribution regularization. All other settings were the same as those in the previous subsection. The results are shown in Table~\ref{tab:imbalanced}. Observe that:

\begin{table}[htpb]     \centering
\caption{Average AUCs  (\%) on the three MI datasets under test-time class-imbalance. The best AUCs are marked in bold, and the second best by an underline.}  \label{tab:imbalanced}
    \begin{tabular}{c|c|c|c}   \toprule
        Approach & BNCI2014001 & BNCI2014002 & BNCI2015001 \\                \midrule
        EEGNet & 78.85$_{\pm1.23}$ & 80.01$_{\pm1.05}$ & 81.26$_{\pm0.32}$ \\
        MCC & 72.94$_{\pm2.44}$ & 77.32$_{\pm2.34}$  & 78.40$_{\pm1.70}$  \\
        SHOT-IM & 74.84$_{\pm1.00}$ & 78.48$_{\pm0.92}$ & 78.60$_{\pm0.89}$ \\
        BN-adapt & 79.14$_{\pm1.35}$ & 79.86$_{\pm0.80}$ & 80.84$_{\pm1.01}$ \\
        PL & 79.93$_{\pm1.05}$ & 82.24$_{\pm1.50}$ & 83.08$_{\pm1.24}$ \\
        SAR & 79.13$_{\pm0.77}$ & 83.44$_{\pm1.72}$ & 82.27$_{\pm0.49}$ \\
        ISFDA (IM) & 79.58$_{\pm0.79}$ & 83.38$_{\pm0.99}$ & 82.75$_{\pm0.44}$ \\
        DELTA (IM) & 79.66$_{\pm1.00}$ & 83.38$_{\pm0.94}$ & 82.89$_{\pm0.48}$ \\
        T-TIME & \underline{81.65}$_{\pm1.39}$ & \underline{85.28}$_{\pm1.04}$ & \underline{84.02}$_{\pm1.17}$ \\
        T-TIME (5) & \textbf{82.72}$_{\pm1.27}$ & \textbf{86.01}$_{\pm0.79}$ & \textbf{84.68}$_{\pm0.52}$ \\
        \bottomrule
    \end{tabular}
\end{table}

\begin{enumerate}
\item Without explicitly considering test-time class-imbalance, MCC, the best-performing offline TL approach in the previous subsection, had the worst performance.

\item Our proposed T-TIME outperformed all approaches, including ISFDA and DELTA, which specifically consider test-time class-imbalance.
\end{enumerate}

\subsection{Different Ensemble Strategies}\label{sect:ensemble}

\begin{figure}[htpb]\centering
\subfigure[]{\includegraphics[width=.9\linewidth,clip]{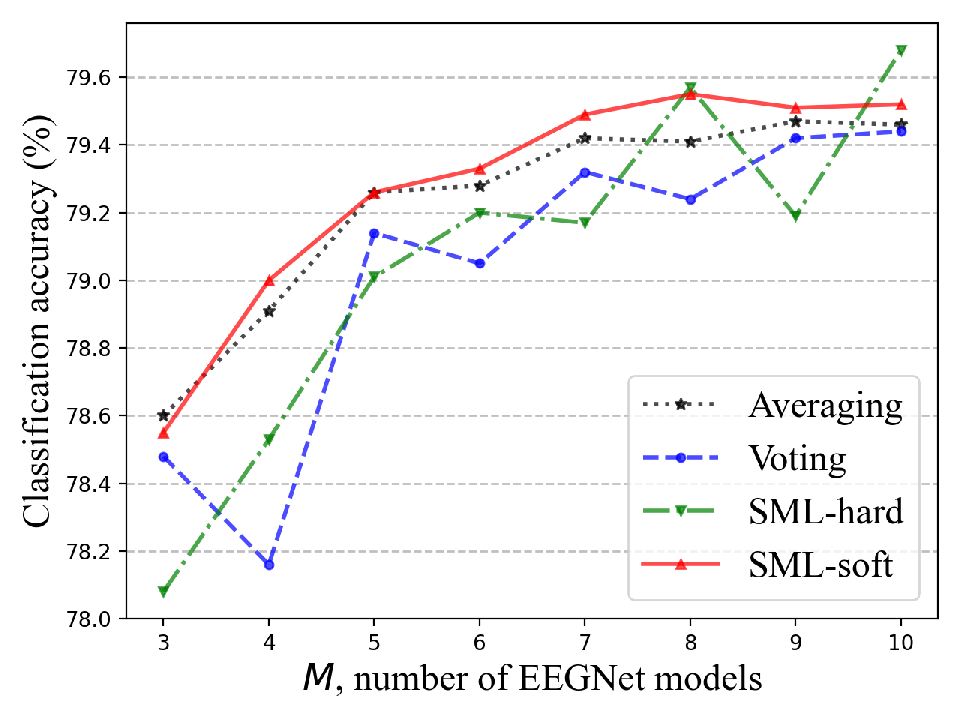}}
\subfigure[]{\includegraphics[width=.9\linewidth,clip]{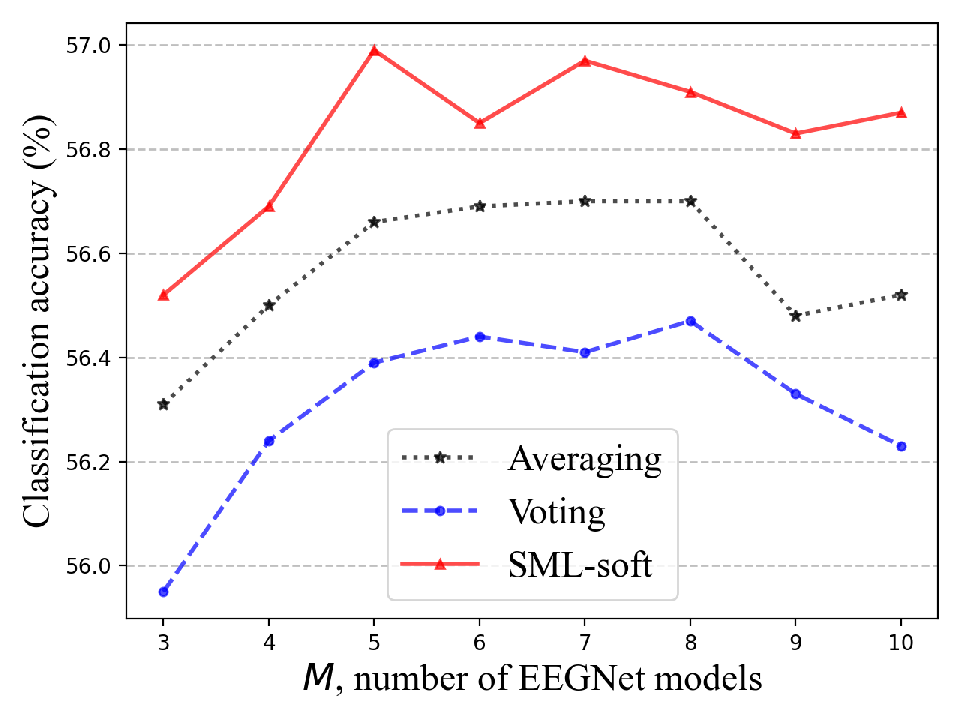}}
\caption{Performance of different ensemble strategies on BNCI2014001 as the number of EEGNet base models varies. (a) binary classification (left/right hand); and, (b) 4-class classification.} \label{fig:ensemble}
\end{figure}

A comparison of different ensemble strategies is shown in Fig.~\ref{fig:ensemble}. With different random seeds and training batch splits, multiple EEGNet source models were trained and updated with T-TIME. Averaging and Voting are classic ensemble strategies. SML-hard and SML-soft used binary predictions and continuous prediction probabilities, respectively (SML-soft is used in our proposed T-TIME). All strategies used identical source models and all experiments were repeated 10 times.

Fig.~\ref{fig:ensemble} shows that Averaging always outperformed Voting, and also generally outperformed SML-hard in binary classification. However, our proposed SML-soft almost always achieved the best performance, and it can be used for both binary and multi-class classifications.

\subsection{Ablation and Parameter Sensitivity Analysis}\label{sect:ablation}

Ablation analysis was conducted to check if each strategy used in T-TIME was effective and necessary. For target model update, the three main components were $\mathcal{L}_{CEM}$ (CEM), $\mathcal{L}_{MDR}$ (MDR), and temperature rescaling (TR) with factor $T=2$. The results are given in Table~\ref{tab:ablation}, which shows that every strategy improved the performance, and using all three together achieved the best performance.

\begin{table}[htbp]  \centering
\addtolength{\tabcolsep}{-4pt}
    \caption{Ablation study results (\%). The best accuracies are marked in bold.} \label{tab:ablation}
    \begin{tabular}{c|c|c|cccc}
    \hline
       \multicolumn{3}{c|}{Strategy} & \multirow{2}{*}{BNCI2014001}  & \multirow{2}{*}{BNCI2014002} & \multirow{2}{*}{BNCI2015001} & \multirow{2}{*}{Avg.} \\ \cline{1-3}
        CEM & MDR & TR & ~ & ~ & ~ & ~ \\ \hline
           $\times$ & $\times$ & $\times$ & 73.52$_{\pm1.14}$ & 72.61$_{\pm0.53}$ & 71.68$_{\pm0.72}$ & 72.60 \\
        \checkmark & $\times$ & $\times$ & 73.90$_{\pm0.95}$ & 76.44$_{\pm0.71}$ & 75.25$_{\pm1.21}$ & 75.20 \\
        $\times$ & \checkmark & $\times$ & 73.72$_{\pm0.79}$ & 74.19$_{\pm0.56}$ & 74.68$_{\pm1.01}$ & 74.20 \\
        $\times$ & $\times$ & \checkmark & / & / & / & / \\
         $\times$ & \checkmark & \checkmark & 73.72$_{\pm0.91}$ & 74.18$_{\pm0.38}$ & 74.19$_{\pm0.89}$ & 74.03 \\
         \checkmark & $\times$ & \checkmark & 74.44$_{\pm0.77}$ & 77.16$_{\pm0.51}$ & 74.84$_{\pm1.26}$ & 75.48 \\
        \checkmark & \checkmark & $\times$ & 76.17$_{\pm1.18}$ & 77.43$_{\pm0.25}$ & 77.27$_{\pm1.47}$ & 76.96 \\
        \checkmark & \checkmark & \checkmark & \textbf{76.44}$_{\pm0.55}$ & \textbf{78.26}$_{\pm0.50}$ & \textbf{77.75}$_{\pm0.67}$ & \textbf{77.48} \\ \hline
    \end{tabular}
\end{table}

\begin{figure}[htpb]
\subfigure[]{\label{fig:sensitivity-T}   \includegraphics[width=.9\linewidth,clip]{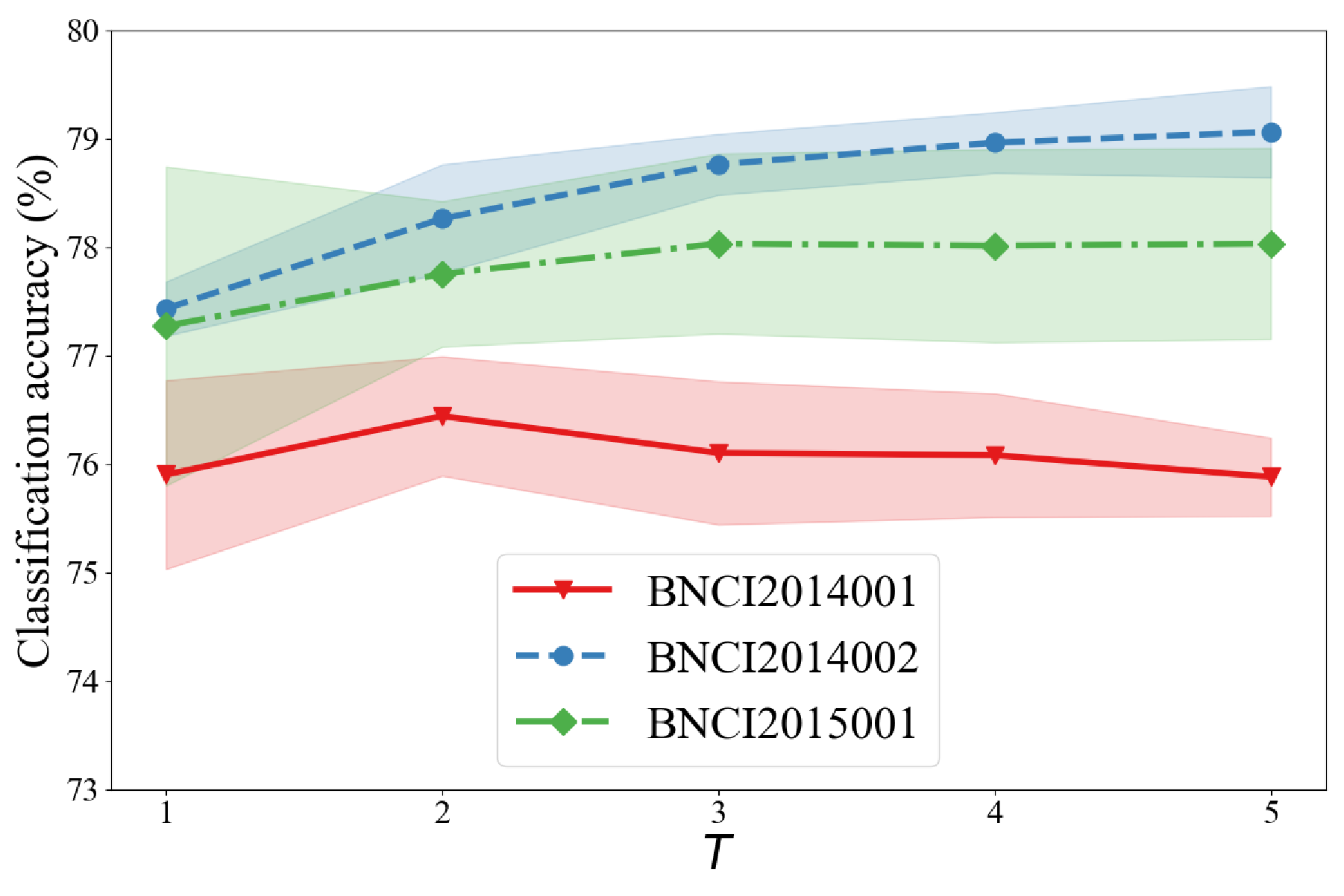}}
\subfigure[]{\label{fig:sensitivity-tau}   \includegraphics[width=.9\linewidth,clip]{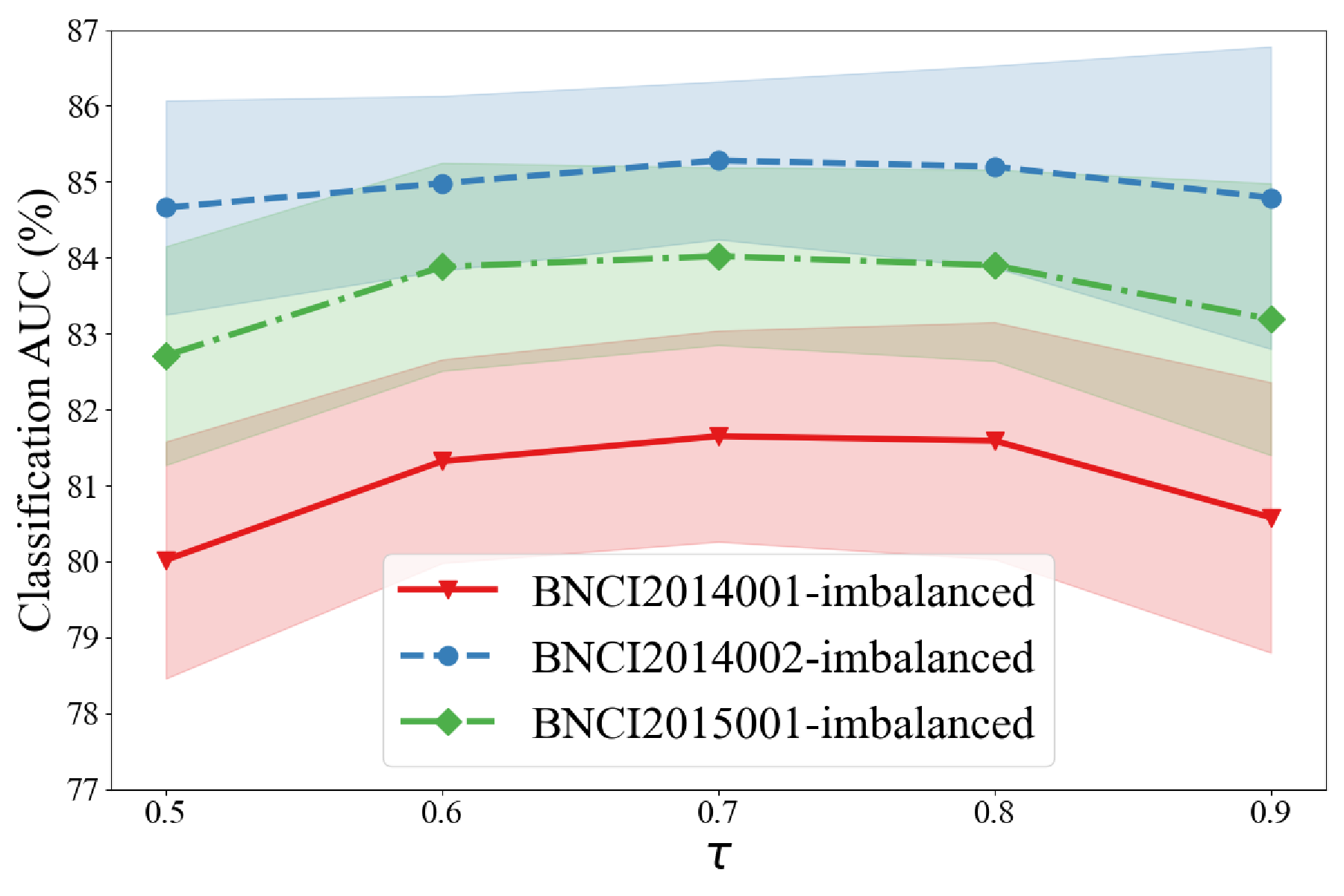}}
\caption{Performance of T-TIME w.r.t. different hyperparameter values. (a) Temperature scaling factor $T$; and, (b) pseudo-labeling confidence threshold $\tau$.} \label{fig:sensitivity}
\end{figure}

Parameter sensitivity analysis was also conducted to validate that a wide range of hyperparameter values in T-TIME could be used to obtain satisfactory performance. Specifically, the two main hyper-parameters, the temperature rescaling factor $T$ and the pseudo-labeling confidence threshold $\tau$, were studied. The results are shown in Fig.~\ref{fig:sensitivity}. $T\in\{2,3,4,5\}$ and $\tau\in[0.6, 0.8]$ worked well on all three datasets.

\subsection{Extension to Continual TTA}\label{sect:continual}

It is well-known that EEG signals are non-stationary, and EEG responses to the same stimulus vary in different sessions from even the same subject~\cite{Liyanage2013}. Thus, it is interesting to investigate continual TTA in unsupervised cross-subject BCIs: after the source model is adapted to the test subject's first session, can it be applied to following sessions from the same test subject?

Four approaches were compared in our experiments:
\begin{enumerate}
\item Source: a model trained in the source domain was directly tested on the second session in the target domain.
\item TTA1: a model trained in the source domain was adapted to the first session in the target domain, and then tested directly on the second session of the target domain.
\item TTA2: a model trained in the source domain was adapted to and tested on the second session in the target domain.
\item TTA1+2: a model trained in the source domain was adapted to the first session in the target domain, and then further adapted to and tested on the second session of the target domain.
\end{enumerate}
The results are given in Table~\ref{tab:continual}, which shows that:

\begin{table}[htpb]    \centering
\caption{Classification accuracies  (\%) on the second session of the three MI datasets. }  \label{tab:continual}
    \begin{tabular}{c|c|c|c}  \toprule
        Approach & BNCI2014001 & BNCI2014002 & BNCI2015001 \\
        \midrule
        Source& 73.78$_{\pm1.85}$ & 74.24$_{\pm1.26}$ & 72.52$_{\pm0.98}$ \\
       TTA1& 75.21$_{\pm1.01}$ & \underline{80.93}$_{\pm1.15}$ & 76.85$_{\pm0.54}$ \\
        TTA2 & \textbf{77.41}$_{\pm1.23}$ & 77.67$_{\pm0.51}$ & \underline{77.19}$_{\pm0.47}$ \\
        TTA1+2 & \underline{77.08}$_{\pm0.85}$ & \textbf{81.60}$_{\pm0.92}$ & \textbf{77.60}$_{\pm0.49}$ \\
        \bottomrule
    \end{tabular}
\end{table}

\begin{enumerate}
\item TTA1 outperformed Source, indicating that the two sessions from the same target subject indeed had some similarity, thus knowledge learned from the first session helped classify the second session.

\item TTA2 outperformed Source, which is intuitive and consistent with the observations in previous subsections.

\item TTA1+2 generally achieved the best performance, suggesting the necessity to always adapt to the new session, and it is beneficial to make use of more history data from the same subject.
\end{enumerate}

\subsection{Computational Cost}\label{sect:comptime}

\begin{figure}[htpb]
\includegraphics[width=\linewidth,clip]{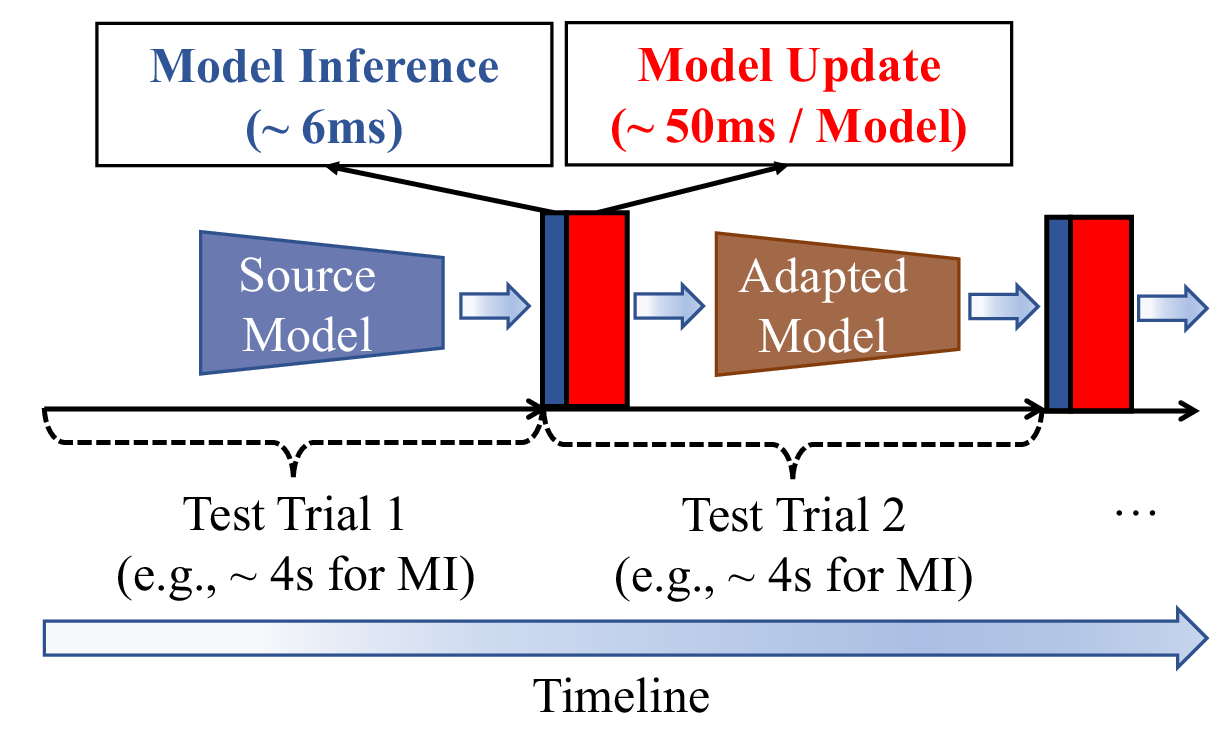}
\caption{Two portions of computation time of T-TIME. The computation time of model inference is the prediction delay, whereas model update can be performed before the next complete test arrives, which usually take seconds.} \label{fig:comptime}
\end{figure}

The computation time of T-TIME on an Intel Core i5 CPU was recorded. Pre-inference target label prediction consists of IEA and SML, which took about 5.3 ms and 0.3 ms on average, respectively. Post-inference target model update took about 34/61/55 ms on average (worst case 63/89/87 ms) for a single EEGNet model on the three datasets. The model update can be carried out in the background until the next test trial finishes (which usually take seconds) and queries for model inference again, as illustrated in Fig.~\ref{fig:comptime}. The specific number of models in the ensemble can vary based on the between-trial time of the specific BCI application. Therefore, T-TIME can cope well with real-world BCI applications.

\section{Conclusions}\label{sect:conclusions}

This paper has proposed T-TIME for unsupervised and online TTA to a new BCI user, making plug-and-play EEG-based BCIs possible. T-TIME initializes multiple classifiers from the aligned source data. When an unlabeled test EEG trial arrives, T-TIME first predicts its labels using spectral meta-learner, and then updates each classifier by conditional entropy minimization and adaptive marginal distribution regularization. Extensive experiments on three public MI-based BCI datasets demonstrated that T-TIME outperformed about 20 classical and state-of-the-art TL approaches. To our knowledge, this is the first study on TTA for calibration-free EEG-based BCIs.

The following directions will be considered in our future research:
\begin{enumerate}
\item We have only considered MI-based BCIs. It is interesting to study whether other BCI paradigms, e.g., event-related potentials and affective BCIs, are also applicable.
\item This paper assumes the source subject is willing to provide his/her classifier to the target subject. For privacy protection, the source subject may encapsulate his/her model as an API, instead of sharing all the details. Privacy-preserving TTA needs to be investigated.
\item In real-world applications of MI-based BCIs, the classifier may not explicitly know the start and end time of each MI trial. How to perform TTA in this challenging situation is another interesting research problem.
\end{enumerate}


\end{document}